\documentclass{aa}
\usepackage[varg]{txfonts}
\usepackage{graphicx}
\usepackage{url}

\title{DOOp, an automated wrapper for DAOSPEC}

\author{
Tristan Cantat-Gaudin\inst{\ref{UniPD},\ref{OAPD}}
\and Paolo Donati\inst{\ref{UniBO},\ref{OABO}}
\and Elena Pancino\inst{\ref{OABO}}
\and Angela Bragaglia\inst{\ref{OABO},\ref{ASI}}
\and Antonella Vallenari\inst{\ref{OAPD}}
\and Eileen D.~Friel\inst{\ref{Indiana}}
\and Rosanna Sordo\inst{\ref{OAPD}}
\and Heather R.~Jacobson\inst{\ref{MIT}}
\and Laura Magrini\inst{\ref{OAAr}}
}

\institute{Dipartimento di Fisica e Astronomia, Universit\`a di Padova, vicolo Osservatorio 3, 35122 Padova, Italy\label{UniPD}
\and INAF-Osservatorio Astronomico di Padova, vicolo Osservatorio 5, 35122 Padova, Italy\label{OAPD}
\and Dipartimento di Fisica e Astronomia, Università di Bologna, Viale Berti Pichat, 6/2 - 40127 Bologna, Italy\label{UniBO}
\and INAF-Osservatorio Astronomico di Bologna, via Ranzani 1, 40127 Bologna, Italy\label{OABO}
\and ASI Science Data Center, I-00044 Frascati, Italy\label{ASI}
\and Department of Astronomy, Indiana University, Bloomington, IN 47405, USA\label{Indiana}
\and Massachusetts Institute of Technology and Kavli Institute for Astrophysics and Space Research, 77 Massachusetts Avenue, Cambridge, MA 02139, USA\label{MIT}
\and INAF-Osservatorio Astrofisico di Arcetri, Largo Enrico Fermi 5, 50125 Firenze, Italy\label{OAAr}
}

\begin{document}

\date{Received date / Accepted date }

\abstract
{Large spectroscopic surveys such as the Gaia-ESO Survey produce huge quantities of data. Automatic tools are necessary in order to efficiently handle this material. The measurement of equivalent widths in stellar spectra is traditionally done by hand or with semi-automatic procedures that are time-consuming and not very robust with respect to the repeatability of the results.}
{The program DAOSPEC is a tool that provides consistent measurements of equivalent widths in stellar spectra while requiring a minimum of user intervention. However, it is not optimised to deal with large batches of spectra, as some parameters still need to be modified and checked by the user. Exploiting the versatility and portability of BASH, we have built a pipeline called DAOSPEC Option Optimiser (DOOp) automating the procedure of equivalent widths measurement with DAOSPEC.}
{DOOp is organised in different modules that run one after the other to perform specific tasks, taking care of the optimisation of the parameters needed to provide the final equivalent widths, and providing log files to ensure better control over the procedure.}
{In this paper, making use of synthetic and observed spectra, we compare the performance of DOOp with other methods, including DAOSPEC used manually. The measurements made by DOOp are identical to the ones produced by DAOSPEC when used manually, while requiring less user intervention, which is especially convenient when dealing with a large quantity of spectra. Like DAOSPEC, DOOp shows its best performance on high-resolution spectra (R>20\,000) and high signal-to-noise ratio (S/N>30), with uncertainties ranging from 6\,m$\mbox{\AA}$ to 2\,m$\mbox{\AA}$. The only subjective parameter that remains is the choice of the normalisation, as the user still has to make a choice on the order of the polynomial used for the continuum fitting. As a test, we use the equivalent widths measured by DOOp to re-derive the stellar parameters of four well-studied stars.}
{}


\maketitle

\section[]{Introduction}\label{intro}
We are presently living in an era of large astronomical surveys that are delivering an unprecedented amount of information, like the current APOGEE \citep[Apache Point Observatory Galactic Evolution Experiment,][]{apogee} and RAVE \citep[Radial Velocity Experiment,][]{rave} and those to come in the near future such as GALAH \citep[Galactic Archaeology with HERMES,][]{galah}, 4MOST \citep[4-metre Multi-Object Spectroscopic Telescope,][]{4most} or MOONS \citep[Multi-Object Optical and Near-infrared Spectrograph,][]{moons}. The Gaia ESO Survey \citep[hereafter GES, see][]{ges12} is a public spectroscopic survey that started at the end of 2011, carried out on FLAMES at Very Large Telescope, targeting more than 10$^5$ stars over the course of five years, systematically covering all major components of the Milky Way, from ancient halo stars to star forming regions, providing the first homogeneous overview of the distributions of kinematics and detailed elemental abundances. The data analysis of the GES is a complex task carried out by different groups. When dealing with a huge quantity of astronomical data, it is essential to have tools that economically process large amounts of information and produce repeatable results. 

Some automatic spectrum analysis procedures rely on the minimisation of the $\chi^{2}$ difference between the observed spectrum and a set of synthetic ones \citep[for instance SME,][]{sme}, or the projection of the spectrum on a vector basis constructed from theoretical spectra \citep[MATISSE]{matisse}. Other procedures are based on the classical method consisting in measuring equivalent widths (EWs) that can be analysed with codes like MOOG \citep{sneden12} or WIDTH9 \citep{kurucz05}. Equivalent widths analysis is widely used, and codes like FAMA \citep{fama} or GALA \citep{gala} were developed recently.

We present here a tool developed to measure the EWs of a large number of spectra in a fully automatic way. This tool, called DAOSPEC Option Optimiser pipeline (DOOp), uses DAOSPEC \citep[][hereafter SP08]{dao08} to measure the EWs and optimises its key parameters in order to make the measurements as robust and repeatable as possible. The aim of this paper is to describe the main characteristics of DOOp and to show the results that can be achieved when combining DOOp with an analysis code such as FAMA, or any other abundance analysis program based on EW measurements.

DOOp, along with a user guide, is available to the community via its webpage: \url{http://web.oapd.inaf.it/GaiaESO/DOOp}

\section{DAOSPEC in a nutshell} \label{sec:dao}
The DAOSPEC code is fully described in SP08, and we will simply note its main characteristics and underline a few points that are important for the best use of DOOp: DAOSPEC is an automated Fortran program to
measure EWs of absorption lines in high-resolution (typically, higher than
20\,000) and high signal to noise (S/N higher than 30) spectra of stellar atmospheres. The measured lines are matched with a user-provided line list.

The code employs a fixed full width at half maximum (or scaled with wavelength for echelle spectra) to
facilitate deblending, and estimates the continuum with Legendre
polynomials after all the fitted lines are removed from the spectrum.
These two characteristics are not present in codes like SPECTRE \citep{fitzpatrick87}, ARES \citep{ares}, or EWDET \citep{ewdet}, that all leave the full width at half maximum as a free parameter for each line, as is commonly done when measuring EWs with the IRAF task {\em splot}. This makes DAOSPEC especially useful on crowded spectra.

One important feature of DAOSPEC (see Fig.~4 in SP08) is that the continuum on which the EW fits are based is not the {\em
true continuum} of the spectrum (i.e., the continuous star emission
after all the lines are excluded), but an {\em effective continuum},
which is the true continuum depressed by a statistical estimate of the
contaminating lines (the unresolved or undetected ones, producing a sort
of line blanketing). This greatly improves the estimate of the unblended
EW of each line in crowded spectra, as demonstrated in Sect.~3.2.1
of SP08, but it is often perceived as being too low by those who are used to employing traditional
interactive methods for the continuum fitting procedure. The discrepancy
between the true continuum and the effective continuum increases with line
crowding (i.e., spectrum metallicity, especially for giants) and with
decreasing S/N or resolution of the spectra.

Full width at half maximum (FWHM) and continuum placement are strictly
correlated: if the continuum level of a spectrum is altered, so is the
FWHM of each line. This is why the three most important parameters for a
successful use of DAOSPEC are: (1) the FWHM estimate;
(2) the continuum placement; and (3) the residual core flux parameter,
which is the flux at the core of saturated lines, expressed in percent of the local continuum level; 
therefore, DOOp is designed to provide the best fine-tuning of these three
parameters.

Another characteristic of DAOSPEC, which is uncommon for EW measurement
programs (both interactive and automated), is that it provides a number of
quality estimates of each EW measurement. These are the formal fitting
errors in the single line, the quality parameter Q of each single line (a
comparison of the local residuals around each line with the average
residuals of the whole spectrum), and the average residuals, expressed as
a percentage, over the whole spectrum. The codes mentioned before do not, to our knowledge, provide an uncertainty on the fit of individual lines (except for EWDET), and none of them performs a global estimate of the quality of fit.
These features of DAOSPEC are used
by DOOp to estimate the effects of a systematically incorrect
continuum placement, for example.

The DAOSPEC code relies on statistical evaluation to consistently estimate the FWHM of the lines and place the effective continuum across the whole spectral range, which means that it performs better on wider ranges than on smaller ones. When dealing with spectra from an echelle spectrograph that delivers individual orders, it is safer to use a merged spectrum rather than measuring each order separately. This applies of course also to DOOp.

\section[]{DOO pipeline}\label{sec_description}
The DOOp code is an algorithm which optimises the parameters of DAOSPEC in order to get the best measurements of EWs. The fine tuning of the parameters is obtained through a fully automatic and iterative procedure and is tailored to the intrinsic characteristic of the spectrum that is going to be analysed. This procedure is performed by different scripts written in BASH\footnote{BASH is a Unix shell, a free software in common with all the operating systems based on UNIX and Linux.} and IRAF\footnote{IRAF is distributed by the National Optical Astronomical Observatory which is operated by the Association of Universities for Researches in Astronomy, under cooperative agreement with the National Science Foundation.} built around DAOSPEC.

The DAOSPEC parameters on which DOOp focuses are the following:

\begin{itemize}
\renewcommand{\labelitemi}{$\bullet$}
\item short wavelength limit (SH)
\item long wavelength limit (LO)
\item minimum radial velocity (MI)
\item maximum radial velocity (MA)
\item residual core flux (RE)
\item FWHM (FW)
\end{itemize}

An exhaustive description of these parameters can be found in SP08 or in the DAOSPEC manual publicly available\footnote{\url{www3.cadc-ccda.hia-iha.nrc-cnrc.gc.ca/community/STETSON/daospec} or \url{www.bo.astro.it/~pancino/projects/daospec.html}}. Here we describe briefly their meaning.
The SH and LO parameters specify the spectral range over which DAOSPEC will measure equivalent widths.
The MI and MA parameters set the velocity range in which DAOSPEC is allowed to estimate the radial velocity (RV) of the star. Imposing a restricted range of possible RV through MI and MA reduces the risk of mismatching the lines (for instance in spectra with very few or very broad lines) and helps to find the right value. It also reduces the computation time.
The RE parameter tells the program the residual flux at the core of the deepest line in the spectrum. 
The FW sets the estimate of the resolution of the spectrum in units of pixel.
All the other DAOSPEC parameters are set to default values but some of them must be specified in the input file of DOOp (see Sect.~\ref{sec:DOOpfunctioning}). 

Among these, the most important is the order of the polynomial (OR) used to fit the continuum. This parameter is $not$ optimised by DOOp, and is kept fixed to the value provided by the user. The OR parameter should be chosen with care and in Sect.~\ref{sec:OR} we discuss its importance and its impact on the EW measurement.

\subsection{The choice of continuum order}\label{sec:OR}
The choice of continuum fitting when analysing a stellar spectrum is always of great importance. A good model of the continuum must follow the main large-scale features in a spectrum, and a general rule of thumb is to use a polynomial of an order similar to the number of waves seen in the spectrum. For a large wavelength range, one must use a polynomial of higher order. Different choices of continuum can result in differences in EWs of up to 2\,m{\AA} for some lines, as will be illustrated in Sect.~\ref{doopvsdao} when comparing measurements with literature values obtained with DAOSPEC.

\subsection[]{Basic functioning}\label{sec:DOOpfunctioning}
\begin{figure*}
\begin{center} \includegraphics[scale=0.40]{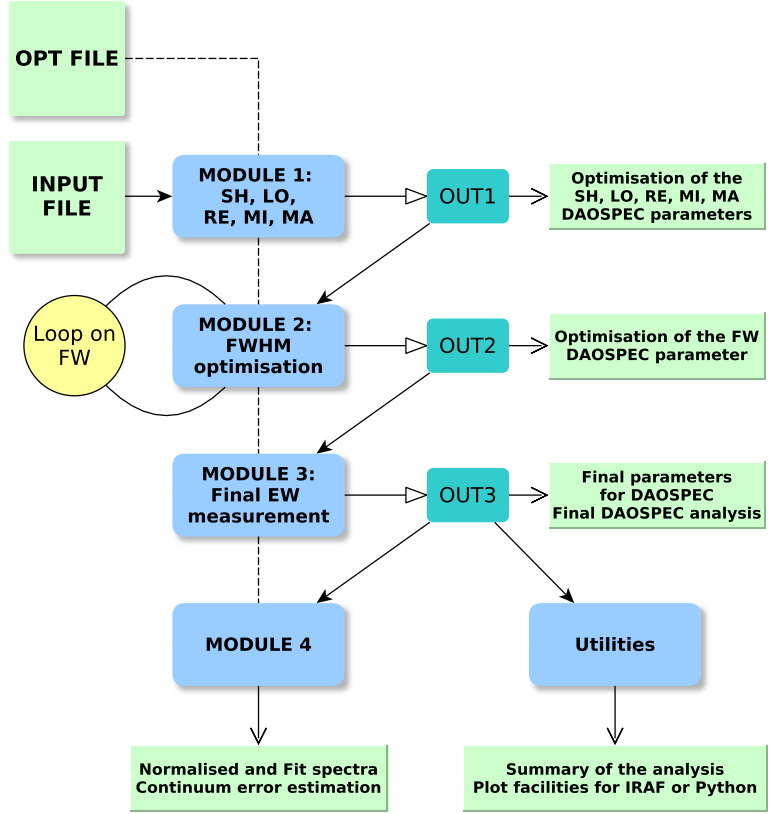} \caption{\label{doop_chart}The tasks performed by DOOp are organised in several modules. The figure shows the dependencies of the modules and their main results.} \end{center}
\end{figure*}
The work flow of the pipeline is summarised in Fig.~\ref{doop_chart}. The pipeline reads two input files. One is needed to set the basic options of the algorithm of the whole pipeline, such as the names of the output files and the convergence parameters for Module~2 (described later in this section). The other input file contains the list of spectra to analyse and a set of six parameters for each of them. The first five parameters are the DAOSPEC parameters OR (order of the polynomial for continuum fitting), FW (first guess of the FWHM), FI (=1 if this FWHM must be kept fixed, =0 if it must be optimised), RE (residual core flux) and RV (radial velocity). If RE and RV are known their value can be given by the user to save computation time, otherwise setting a value of 0 means they will be derived by the pipeline. The last parameter sets which algorithm the pipeline must use to optimise the FWHM. At the moment two possibilities are allowed (see next paragraphs). All the parameters must be explicitly specified. To ensure that DOOp performs well on a list of spectra it is always better to analyse together spectra of the same resolution, that were collected with the same instrument.

\begin{figure}[h]
\begin{center} \resizebox{\hsize}{!}{\includegraphics[scale=0.55]{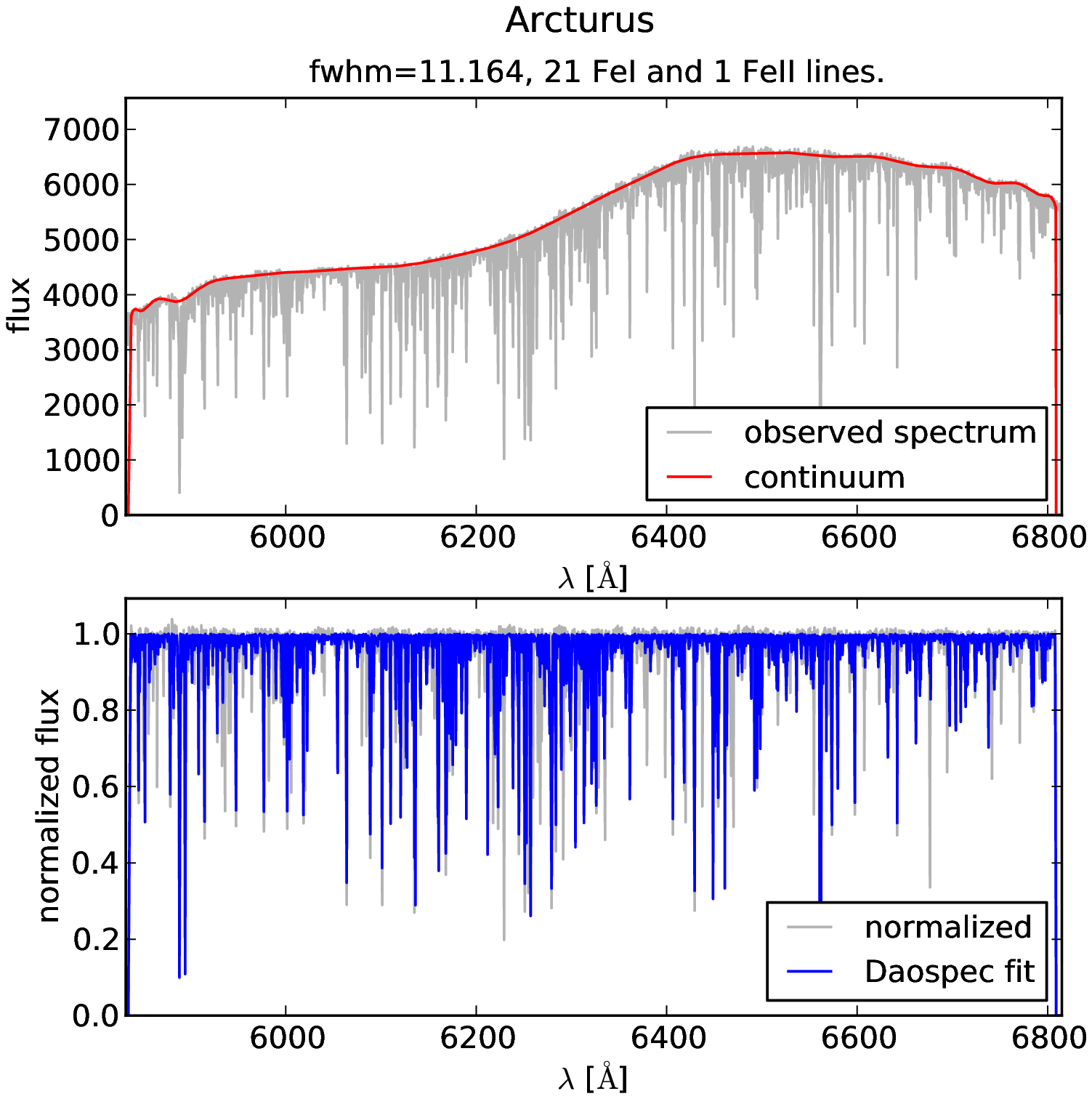}} \caption{\label{arcturusSpec}UVES-POP spectrum of Arcturus seen through the graphical interface of DOOp. \textit{Top:} original spectrum and fitted continuum. The flux is given in arbitrary units, as the instrument response was not corrected for. Information is displayed on the FWHM of the lines and the number of lines that were identified from the line list. \textit{Bottom:} normalised spectrum and fit.} \end{center}
\end{figure}

After these two files are set, DOOp is ready to work. The measurements will be carried by four modules performing different tasks:

\begin{itemize}
\renewcommand{\labelitemi}{$\bullet$}
  \item Module~1: this module provides the SH, LO, RE, MI, and MA parameters, as well as the first EW estimates. The SH, LO, and RE values are determined calling IRAF twice with two different IRAF scripts. The first looks for the starting and ending wavelength of the spectrum avoiding glitches and bad values that may be found at the spectrum borders, setting the two parameters as the first/last wavelength for which the spectrum has a non-negative value. The second looks for the strongest lines in the spectrum (H$\alpha$, H$\beta$, Mg b triplet, CaII triplet) to set properly the RE parameter. If negative values are found, or if none of these strong lines are detected, a default value is imposed. The MI and MA parameters are set accordingly to the input RV: if 0, then a wide range is imposed, otherwise a smaller range is used. These ranges can be defined by the user ($\pm$ 500 km/s and $\pm$ 10 km/s are typically reasonable values). At the end of Module~1 a first EW measurement is performed on the spectra for which DAOSPEC did not encounter computational problems. This measurement is not the best one, as not all the parameters are optimised.

  \item Module~2: this module provides the best FWHM parameter for DAOSPEC. For each analysed spectrum, an initial value of FWHM is needed. If the user wishes to optimise the FWHM (by setting the parameter FI=0), DAOSPEC runs once and Module~2 compares the output FWHM to the input (user-given) value. If a convergence criterion (by default, the output value has to be within 3\% of the input) is not reached, then DAOSPEC runs again, using the output value as a new initial guess. The user can choose how many of the spectra in the list have to reach convergence (in percent). In some cases, the spectra under analysis are known a priori to present the same FWHM but its estimate is made difficult in some spectra (for instance because of different signal-to-noise ratios). In such a case it can be sufficient to reach convergence for only 50 or 70\% of them and use the median value of the FWHM to measure the others (which is done by Module~3). In the most general case of a batch containing spectra of potentially different FWHMs (because the spectra were obtained with different instruments, under different sky conditions or because of rotating stars) it is recommended to require 100\% of the spectra to reach convergence of the FWHM, so that this parameter is estimated in an independant way for each spectrum. If after 30 iterations the FWHM has not converged for some spectra, Module~3 will take care of them.

  \item Module~3: this module determines the FWHM that will be used for the final EW measurements. It computes the median FWHM of all the spectra for which it converged. Depending on the choice of the user, it will use this median value for the spectra that did not converge, or for all the spectra in the batch {\em including those that converged}. This second option may be more suitable if all the spectra are known a priori to have the same FWHM. Of course, the spectra for which the user had required to use a fixed FWHM (by setting FI=1) are measured using the FWHM given by the user. The output of DAOSPEC at the end of Module~3 are {\em the final EW measurements of DOOp}.

  \item Module~4 is designed to perform two different tasks. One is to provide the fit and normalised spectra in FITS format, because DAOSPEC only provides the continuum and residual FITS files. The other task is to provide EW measurements obtained by over- and underestimating the continuum level. The amount by which the continuum will be shifted is proportional to the dispersion of the residuals in the final DAOSPEC fit. The results coming out of these experiments can be used to quantify the error in the EW measurement due to the placement of the continuum. For UVES spectra (R=47\,000), the dispersion of the residuals of the fit ranges typically from 3\% for a S/N of 30, to 1\% or less for a S/N above 100.  Altering the continuum placement by these amounts can lead to differences of 10 and 3 m$\mbox{\AA}$ respectively, although this is certainly a conservative estimate and can depend on other factors such as the metallicity of the star. A description of how the EWs are changed when altering the continuum placement is done by SP08 in their Sect.~3.4.3 and Fig.~2. Module~4 uses IRAF script to obtain the FITS files, while DAOSPEC is called to perform the EW measurements on the spectra with artificially imposed continuum levels. This module works with the output produced by Module~3.

  \item Utilities: together with the main algorithm of the pipeline, small scripts are provided to perform standard operations on the output files obtained by Module~3. They are presently used to produce the input files for the abundance analysis programs GALA and FAMA and to print out a summary of the analysis, including a log of the possible errors. With a provided script, the user can easily visualise and compare the spectra and their corresponding fit (see Fig.~\ref{arcturusSpec}).
\end{itemize}

One advantage of the pipeline is that it can be easily customised. For example, if only the RV measurement is needed, one can set the pipeline to use only Module~1. If, instead, one wants to test the effect of changing the continuum order while keeping all the other parameters fixed, it is also possible. Furthermore, if the user prefers to pre-normalise their spectra using a personal routine, indicating a continuum order of -1 tells DAOSPEC not to perform any normalisation.

\begin{figure}
\begin{center} \resizebox{\hsize}{!}{\includegraphics[scale=0.55]{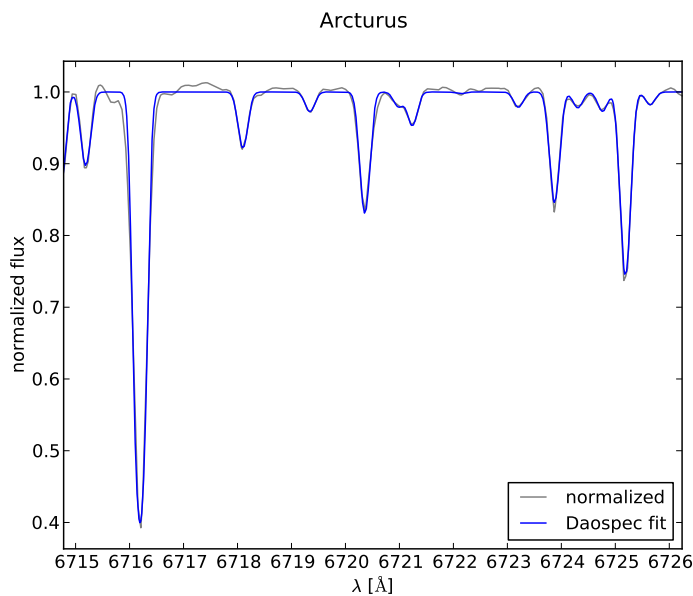}} \caption{\label{arcturusSpecDetail}Detail of the normalised spectrum of Arcturus from Fig.~\ref{arcturusSpec} and corresponding fit.} \end{center}
\end{figure}

Examples of the resulting fits obtained with DOOp on a spectrum of Arcturus can be seen in Figs.~\ref{arcturusSpec} and \ref{arcturusSpecDetail}.

\subsection{Technical aspects}
DOOp exploits the versatility of the BASH shell, which usually comes together with a set of smaller stable and powerful programsl to perform operations on files and data. It uses the BASH language and a small set of these programs to handle all the logical operations needed to accomplish the optimisation of the DAOSPEC parameters. The external programs used by the pipeline are DAOSPEC, IRAF, and Python. The latter is called to display the interactively zoomable plots that allow the quality of the fits to be controlled.

On one hand, a script code is more fragile than monolithic codes because it does not pass any compilation checks and moreover the use of different programs interacting together can be less portable. On the other hand, script languages (Perl, Python, and JavaScript, to name a few well-known examples) are now widely used to make small programs for which the computational speed is not a key requirement. They are particularly intuitive, easy, and fast to use and their popularity often ensures the portability of the code.

From the beginning, DOOp was meant to be used by different groups in different locations and the need for portability was a top priority. It has been tested on different operating systems with different versions of BASH and IRAF, hence we can guarantee a full compatibility with at least the software we could test. It has been used on 32-bit and 64-bit Linux kernels (Ubuntu and Cent OS distributions) and with Mac OS. The BASH versions of the machines used range from the old 3.2.48 (2007) to the newer 4.1.2 (2009). We used two different IRAF versions: 2.14 and 2.16.

\subsection{The advantage of DOOp over using DAOSPEC manually}
The DOOp pipeline provides a robust method of measuring EWs in stellar spectra by limiting as much as possible human intervention in the process and facilitating the control of the results. This makes it especially convenient when dealing with batches of hundreds or thousands of spectra produced during observational campaigns such as the GES, and when different groups work with the same data.

For example, the estimate of the RE parameter usually requires manual measurements for each spectrum, while DOOp performs automatic measurements. It also automatically sets the SH and LO parameters for each spectrum, with great gain of time, and reduces the velocity range in which DAOSPEC looks for the radial velocity when the RV parameter is at least approximately known.

Ensuring that DAOSPEC converges to the best value for the FWHM in the spectrum can be a very lengthy procedure if done manually, even for a single spectrum. We show in Sect.~\ref{sec:synth_fwhm} how the choice of initial FWHM can affect the convergence and the final EW measurements. The DOOp code takes care of this convergence process automatically.

The DAOSPEC code works by fitting gaussians to the absorption lines but does not return any file containing the global fitted spectrum. Instead, it returns a fits file containing the continuum that was fitted, and a second file containing the residuals of the fit. The DOOp code automatically combines these files with the original spectrum to create a file containing the fitted spectrum. It also contains tools for direct plotting of the results (using the popular Python library Matplotlib). Each module prints out a log file containing a summary of its action on each spectrum, thus allowing an efficient control over the whole procedure.

When DAOSPEC is run, the format of the output files matches the line list provided by the user. In its current version, DOOp can automatically convert the output files produced by DAOSPEC to the format needed by FAMA and GALA if the user provides a line list in the right format. Given the structure of DOOp, organised in independent modules, adapting the code to work with a custom format of line list or produce a specific format of output can be easily done. In any case, the output files of DAOSPEC are kept by DOOp, and one can use them exactly as they would when running DAOSPEC manually.

\section[]{Tests on synthetic spectra}\label{sec_synth}
We applied DOOp to the synthetic spectra used by SP08 to explore the boundaries in S/N, resolution, and pixel sampling under which DAOSPEC performs optimally. We did not run the tests on the spectra of resolution 5\,000 and 10\,000, as SP08 show that these resolutions are too low for DAOSPEC to perform well (and too low for a reliable EW analysis).

\subsection{Convergence of the FWHM} \label{sec:synth_fwhm}
\begin{figure*}
\begin{center} \includegraphics[scale=0.6]{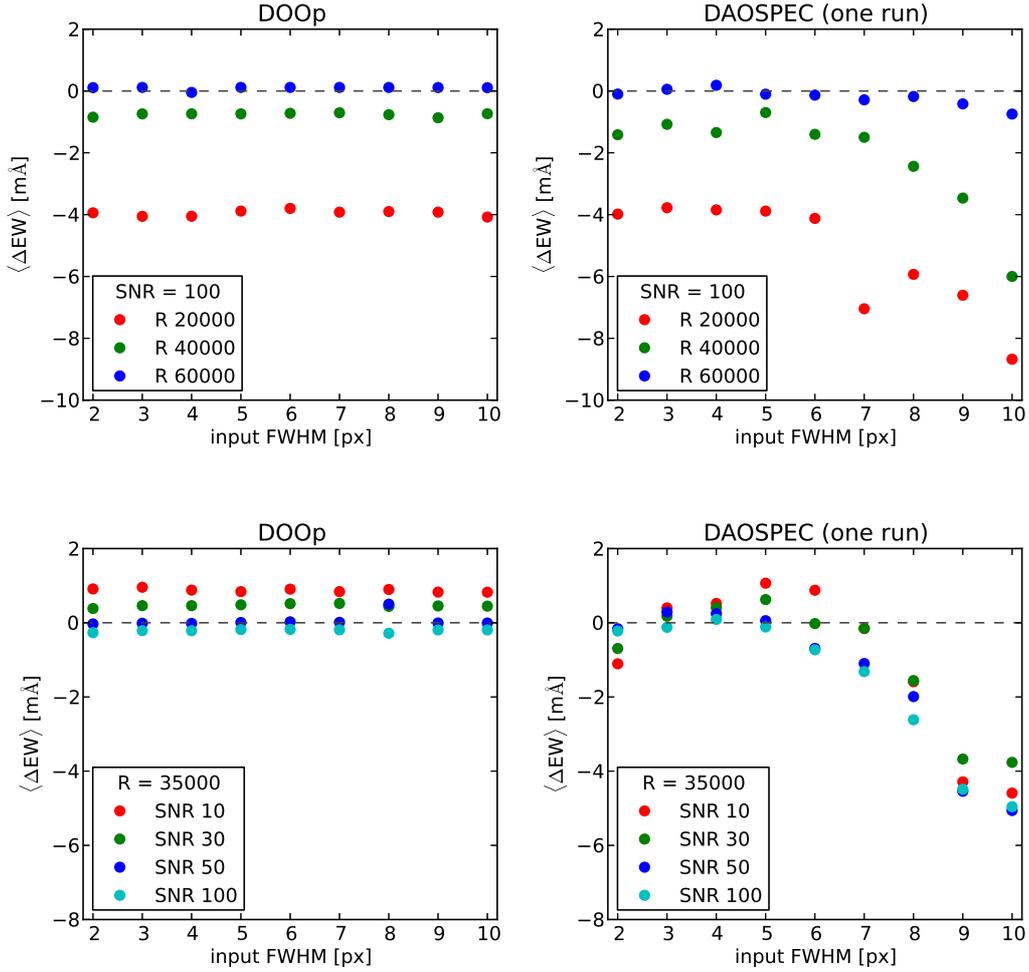} \caption{\label{fwhm_conv}
Difference between our measurements and the true EW for synthetic spectra with a true FWHM of 5\,px, at various resolutions and S/N.
\textit{Top-left:} average difference between the EWs measured with DOOp and the true EWs, for various input values of the FWHM, in synthetic spectra at three different resolutions. 
\textit{Top-right:} same as top-left panel, but for measurements obtained without imposing the convergence of the FWHM.
\textit{Bottom-left:} same as top-left panel, for spectra of four different S/N. 
\textit{Bottom-right:} same as bottom-left, but without imposing the convergennce of the FWHM.} \end{center}
\end{figure*}

We have run tests on the synthetic spectra used in SP08 to compare the EWs recovered after single run of DAOSPEC (which means that no condition on the convergence of the FWHM is imposed) and a run of DOOp (which automatically imposes the convergence of the FWHM), on spectra of various resolutions and signal-to-noise ratios. 
The true FWHM of these synthetic spectra is 5\,px.
The top-right and bottom-right panels of Fig.~\ref{fwhm_conv} show the average difference between the true EWs and the EWs measured with DAOSPEC after one run only, starting from different initial values of the FWHM. The measurements are clearly dependent on the initial FWHM. The top-left and bottom-left panels of the same figure show the difference between the true and measured EWs, after DOOp has imposed the convergence of the FWHM. It is clear that the convergence process carried out by DOOp makes the measurements independent of the input FWHM. This experiment was conducted on spectra at various resolutions and signal-to-noise ratios, showing that DOOp is more stable in all cases.

\begin{figure}
\begin{center} \resizebox{\hsize}{!}{\includegraphics[scale=0.55]{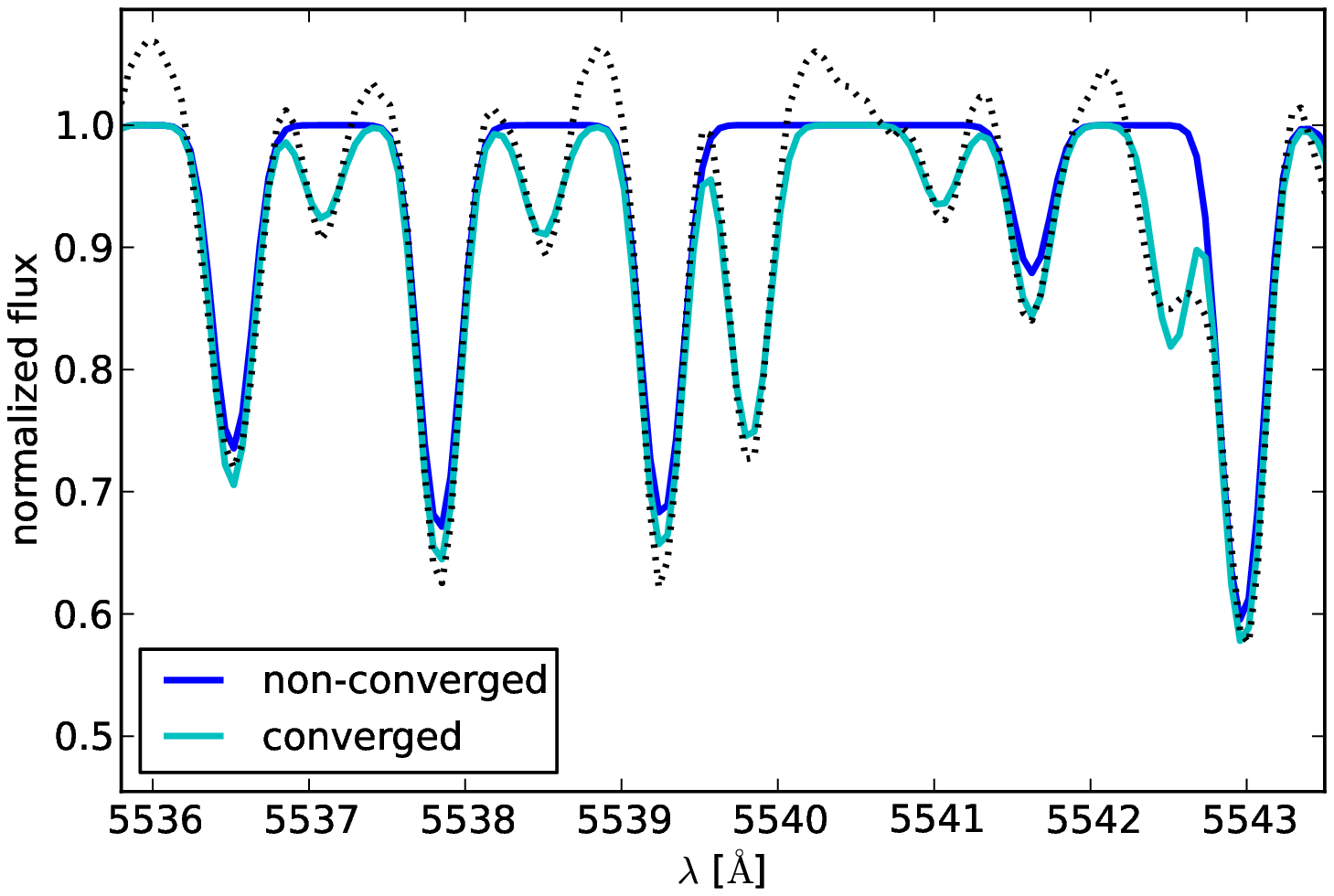}} \caption{\label{lineDet}The dotted line is a synthetic spectrum of resolution R=20\,000. The true value of the FWHM for this spectrum is 5\,px, but for this test we have used 10\,px as a starting point. The blue line is the fit obtained with DAOSPEC run only once. The cyan line is the fit obtained after convergence of the FWHM. Refining the value of the FWHM not only allows for a better fit of the line profiles, but also avoids DAOSPEC missing some lines.} \end{center}
\end{figure}

Ensuring the convergence of the FWHM can also have consequences on the line detection. Starting from a too large value will cause DAOSPEC to ignore some features that it should actually be fitting. Running DAOSPEC again using the newly found FWHM as a new initial value for the convergence helps to find all the lines. Figure~\ref{lineDet} shows two fits obtained with an input FWHM of 10\,px, on a spectrum of resolution R=20\,000 and S/N=100. When DAOSPEC is run only once with a wrong input FWHM it may fail to detect all the lines, but imposing the convergence process improves the line detection.

\subsection{Equivalent widths}
\begin{figure}
\begin{center} \resizebox{\hsize}{!}{\includegraphics[scale=0.6]{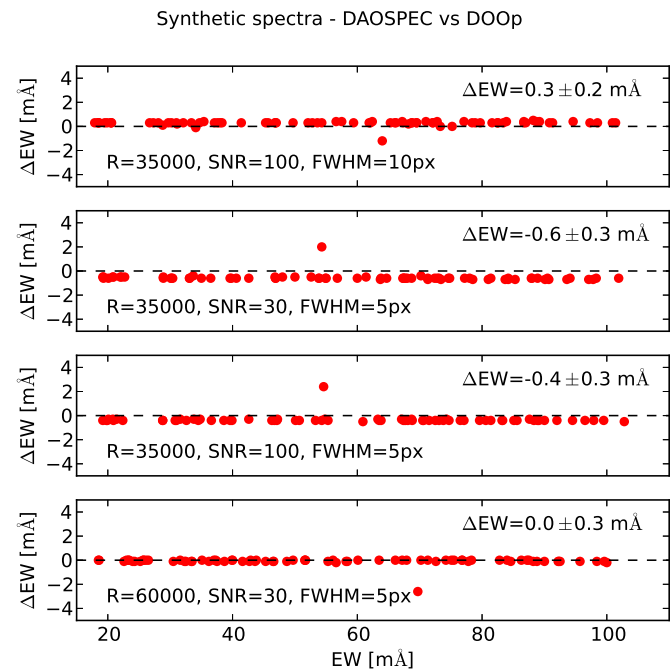}} \caption{\label{synth_onefig}Difference in EW (in the sense DOOp-SP08) for lines in four different synthetic spectra. The results for all the synthetic spectra used in these tests are given in Table~\ref{tab:compsynth}.} \end{center}
\end{figure}

We have compared the measurements of DOOp, with those of SP08, who optimised manually all the parameters (implying multiple runs of DAOSPEC). The spectra used for these tests are the same as in Sect.~\ref{sec:synth_fwhm}, with an additional four spectra of different pixel sampling (true FWHM of 1, 2, 3 and 10\,px). The results of DOOp show excellent agreement with those obtained by SP08. The average differences and the dispersion between both sets of measurements are reported in Table~\ref{tab:compsynth}.
The fit uncertainty is the average uncertainty on the fit of the single lines, that DAOSPEC computes from the least-square fitting procedure (see SP08 for details). The difference and uncertainty decrease with pixel sampling, resolution and signal-to-noise, as is expected. Four of these cases are illustrated in Fig.~\ref{synth_onefig}.

\begin{table}
	\caption{ \label{tab:compsynth} Comparison between the EWs measured by DOOp and by DAOSPEC on synthetic spectra.}
	\small\addtolength{\tabcolsep}{-1pt}
	\begin{tabular}{l l l l l l}
	  \hline
	  \hline
	  R & S/N & FWHM & $\langle \Delta$EW$\rangle$ & r.m.s.   & fit uncertainty\\
	    &     & [px] & [m$\mbox{\AA}$]                   & [m$\mbox{\AA}$] & [m$\mbox{\AA}$] \\
	  \hline
		35\,000 & 100 & 1  &  -0.7  &  0.2 & 5.1\\
		35\,000 & 100 & 2  &  -0.3  &  0.1 & 6.3\\
		35\,000 & 100 & 3  &  -0.4  &  0.1 & 4.6\\
		35\,000 & 100 & 10 &   0.3  &  0.2 & 2.6\\
	  \hline
		20\,000 & 100 & 5 &  -0.3  &  0.4 & 5.6\\
		40\,000 & 100 & 5 &  -0.3  &  0.1 & 3.4\\
		60\,000 & 100 & 5 &   0.0  &  0.3 & 1.8\\
	  \hline
		35\,000 & 10  & 5 &  -0.6  &  0.1 & 4.8\\
		35\,000 & 30  & 5 &  -0.6  &  0.3 & 3.7\\
		35\,000 & 50  & 5 &  -0.6  &  0.1 & 3.6\\
		35\,000 & 100 & 5 &  -0.4  &  0.3 & 3.4\\
	  \hline
	\end{tabular}
\tablefoot{The difference is given in the sense DOOp-SP08.}
\end{table}

\section[]{Comparing the measurements of DOOp with literature}\label{sec_comparisons}
We have checked the measurements obtained with DOOp against already published measurements obtained with other methods, to ensure that our method is able to reproduce the same results, in particular those of DAOSPEC used manually, on real stellar spectra. The stars used in these comparisons are giant, metal-rich stars (-0.5<[Fe/H]<0.1).

\subsection{DOOp vs SPLOT}\label{doopvssplot}
\begin{figure}
\begin{center} \resizebox{\hsize}{!}{\includegraphics[scale=0.5]{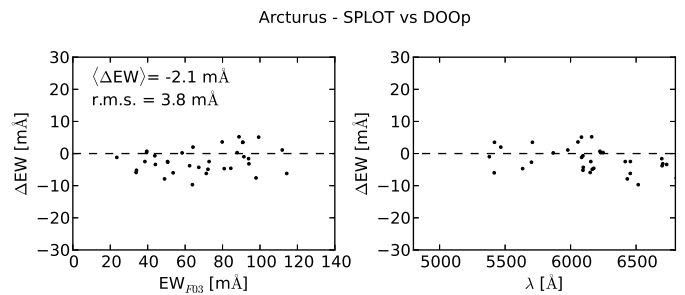}} \caption{\label{compareEWarcturus}Difference in EW (in the sense DOOp-F03) plotted against EW (left panel) and wavelength (right panel) for Arcturus. The average difference is -2.1\,m{\AA}, with an r.m.s. dispersion of 3.8\,m{\AA}.} \end{center}
\end{figure}

We have measured EWs in a spectrum of Arcturus downloaded from the UVES-POP archive\footnote{\url{http://www.eso.org/sci/observing/tools/uvespop.html}} \citep{bagnulo} and degraded from a resolution of 80\,000 to 47\,000, which is the resolution of the UVES-FLAMES spectra.. These EWs were compared with measurements by \citet{friel03}, hereafter F03, on the high resolution spectrum of Arcturus published by \citet{hinkle00}. They have normalised the spectrum using the IRAF CONTINUUM task and measured the EWs using the IRAF task SPLOT. The comparison (Fig.~\ref{compareEWarcturus}) shows good agreement. The DOOp measurements have a general offset of -2.1\,m{\AA}, with an r.m.s. dispersion of 3.8\,m{\AA}. An offset of this magnitude is consistent with the way DAOSPEC sets the continuum level (as discussed in Sect.~\ref{sec:dao} above).

\subsection{DOOp vs SPECTRE}
\begin{figure}
\begin{center} \resizebox{\hsize}{!}{\includegraphics[scale=0.5]{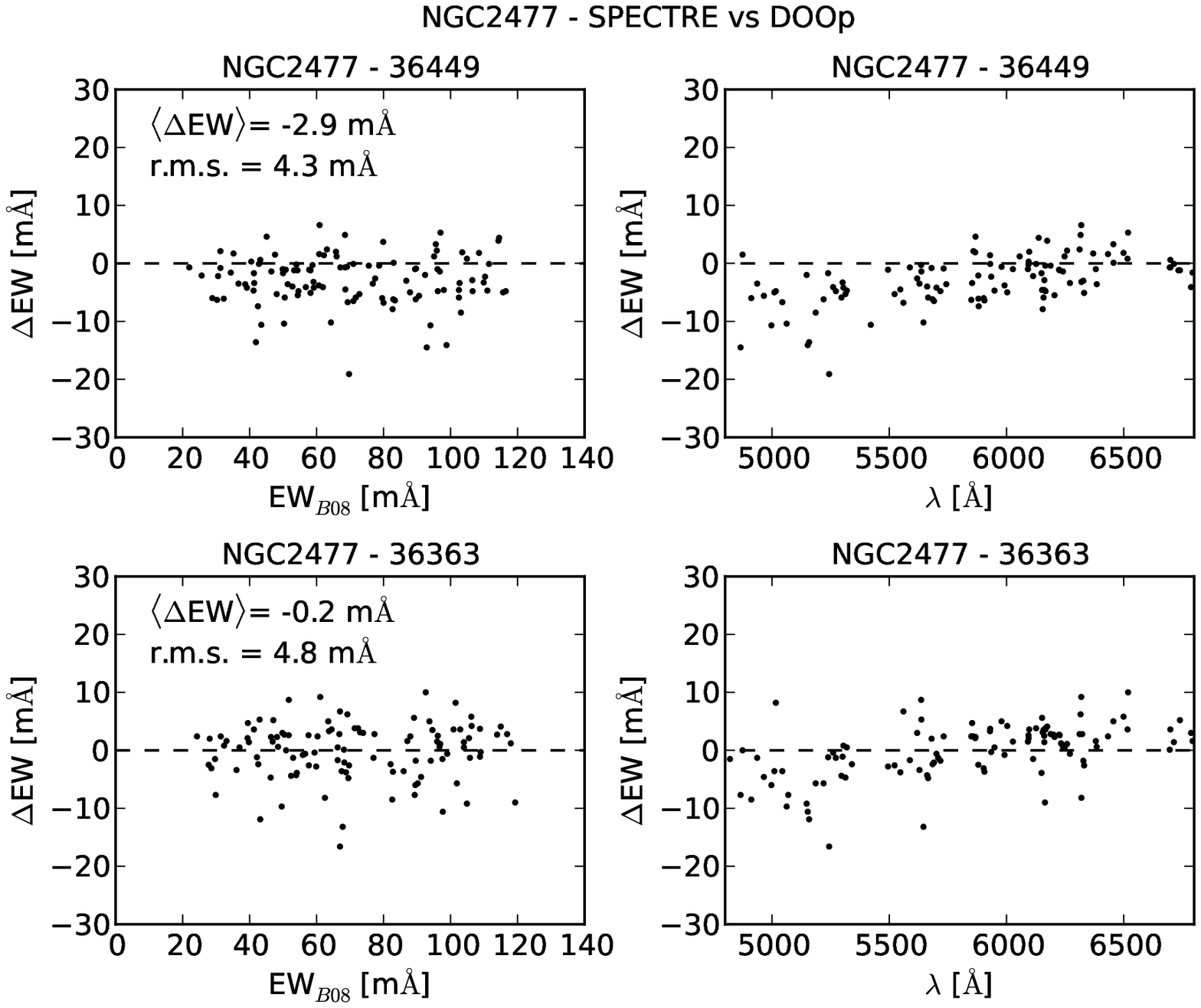}} \caption{\label{compareEWngc2477_2}Difference in EW (in the sense DOOp-B08) plotted against EW (left) and wavelength (right) for two stars of NGC2477.} \end{center}
\end{figure}

We have measured EWs in two stars of NGC~2477 and six stars of Be~29, in UVES spectra previously measured and published by \citet{sest08} and \citet{bragaglia08} (hereafter B08), respectively. These stars are red giants of nearly solar metallicity. The authors have normalised the spectra using the IRAF task CONTINUUM, and fitted Gaussian profiles using the code SPECTRE. We measure in general smaller EWs by 1 to 5~m{\AA}. Again, this can be explained by the behaviour of DAOSPEC in terms of continuum placement.
The results for two stars of NGC~2477 are illustrated in Fig.~\ref{compareEWngc2477_2}. The identifiers of the stars are taken from the ESO Imaging Survey catalogue, as reported in B08. The offset between the measurements of B08 and those of DOOp is inside the range of differences expected when comparing different methods (see for instance the comparisons of SP08 between DAOSPEC and other tools). The slightly different offset observed between the blue and red part of the spectral range comes from the fact that UVES spectra are split between a blue and a red arm, and both ranges were measured independently (by B08 and by us), resulting in a slightly different continuum adjustment.

\subsection{DOOp vs DAOSPEC}\label{doopvsdao}
\begin{figure}
\begin{center} \resizebox{\hsize}{!}{\includegraphics[scale=0.5]{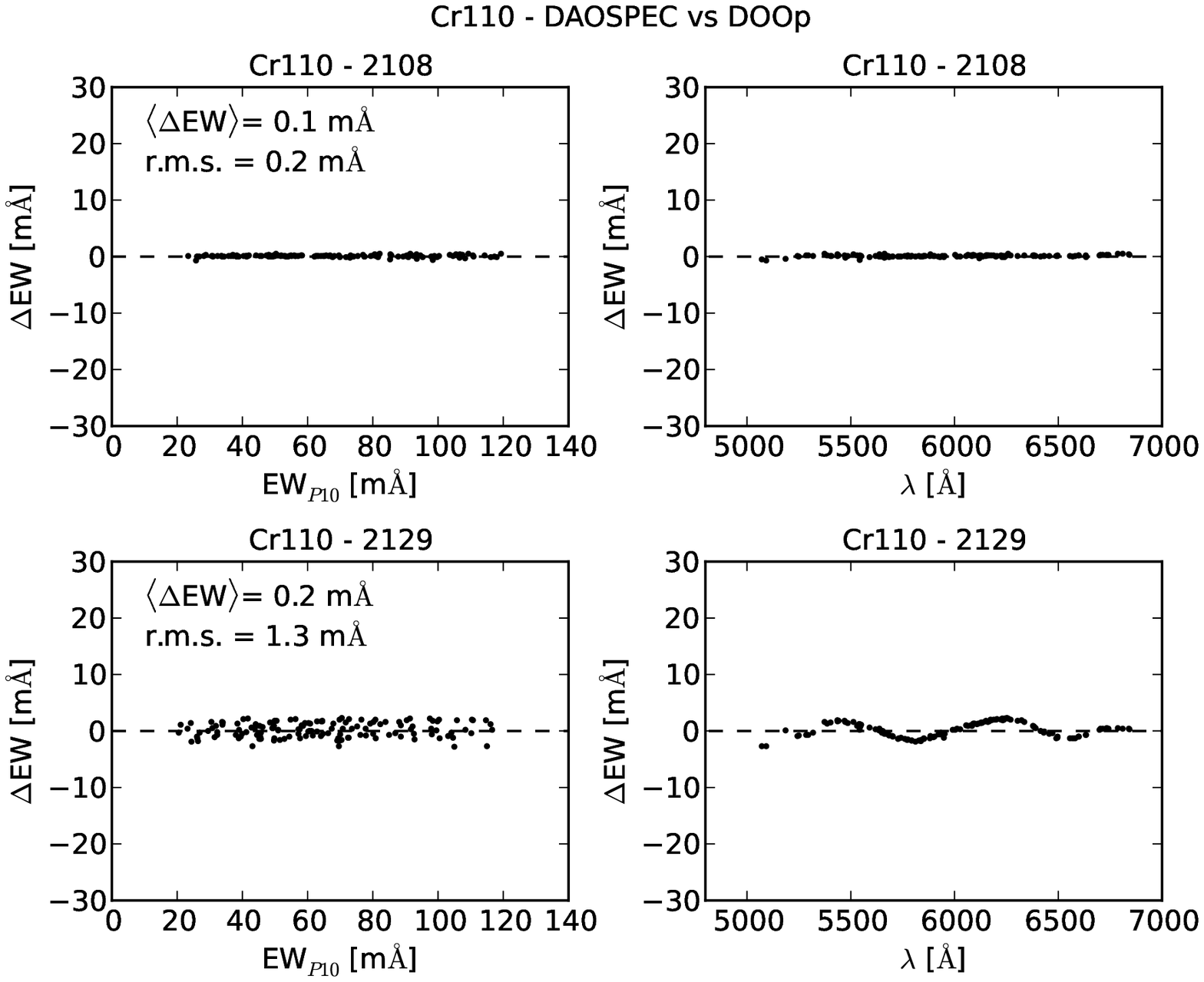}} \caption{\label{compareEWcr110_3}Difference in EW (in the sense DOOp-P10) plotted against EW (left) and wavelength (right) for two stars of Cr110.} \end{center}
\end{figure}

As a final test we have compared our EWs with measurements by \citet{pancino10}, hereafter P10, using DAOSPEC on three stars of Cr~110 and two stars of NGC~2420. These stars are red giants of solar metallicity. The spectra have a resolution of 30\,000 and a S/N of 70. Figure~\ref{compareEWcr110_3} shows the perfect agreement between the two sets of measurements for two stars of Cr~110  \citep[identifiers from][]{Dawson98}. Such a good agreement is expected if the FWHM and residual core flux were carefully set when using DAOSPEC manually, which can be time consuming. The DOOp code does not produce better results than those expected from the most careful use of DAOSPEC, but making the procedure automatic reduces the sources of errors and makes it humanly possible to deal with large numbers of spectra.

However, the critical issue of setting the continuum remains. The refinements of choosing a continuum order remain arbitrary, and two users fitting a continuum of slightly different order on the same spectrum may find slight differences in the measurements of EWs (rms of about 2 m{\AA}). The star 2129 of Fig.~\ref{compareEWcr110_3} is an example of such a case, where small differences varying across the spectral range can be observed.

\section[]{Analysis of benchmark stars}\label{sec_uvespop}
To assess the effect of the different EW measurements on the determination of the stellar parameters, the stars used in Sect.~\ref{sec_comparisons} are not ideal test cases because the authors use different line lists, and using only the lines in common between their lists and ours does not provide sufficient statistics (for instance, we need enough FeII lines in common to derive a reliable gravity).

As a final validation for the whole procedure, we measured the EWs for four well-studied stars, \object{Arcturus}, \object{Procyon}, \object{HD~23249}, and the \object{Sun}. The first three are bright stars (V=-0.04, 0.37, and 3.51, respectively) for which a rich literature is available in the PASTEL database\footnote{\url{http://vizier.u-strasbg.fr/viz-bin/VizieR?-source=B/pastel}}. Their spectra were taken from the UVES-POP archive. These spectra were obtained at the VLT with the UVES instrument at a resolution R$\sim80\,000$ that we degraded to $47\,000$ (which is the nominal resolution of the GES UVES spectra). The list of spectral lines we have used is the GES line list \citep{Heiter13}. The available UVES-POP data cover the optical range, from $3040$ to $10\,400\,${\AA}, of which we have used the ranges $4760 - 5770$ and $5840 - 6840\,${\AA}. These are the ranges covered by most of the GES UVES spectra (corresponding to the setting 580), where the line list is well-defined.

The solar spectrum we analysed was taken from the HARPS archive\footnote{\url{http://archive.eso.org/wdb/wdb/eso/repro/form}}, degraded from a resolution R$\sim$120\,000 to 47\,000. After running DOOp, we passed the output files to FAMA to obtain the atmospheric parameters of these stars.

\begin{figure}
\begin{center} \resizebox{\hsize}{!}{\includegraphics[scale=0.6]{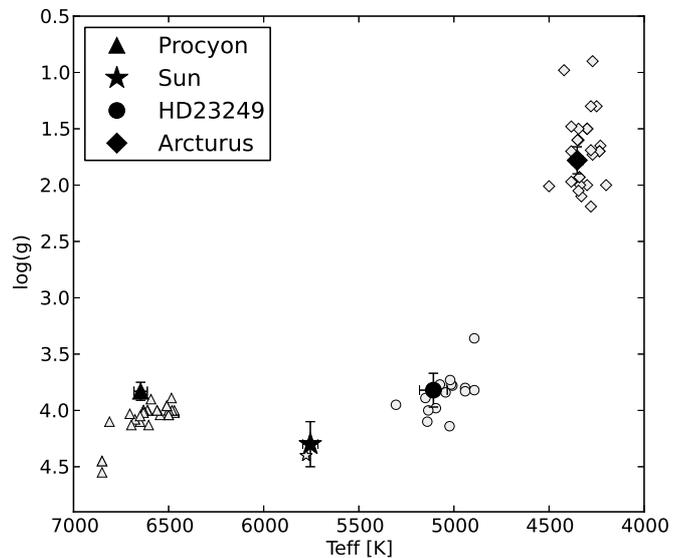}} \caption{\label{3bmTefflogg}T$_{\text{eff}}$ and $\text{log}\,g$ for our four benchmarks. The filled symbols are our results, while the empty symbols are the various values found in the literature.} \end{center}
\end{figure}

\begin{figure}
\begin{center} \resizebox{\hsize}{!}{\includegraphics[scale=0.6]{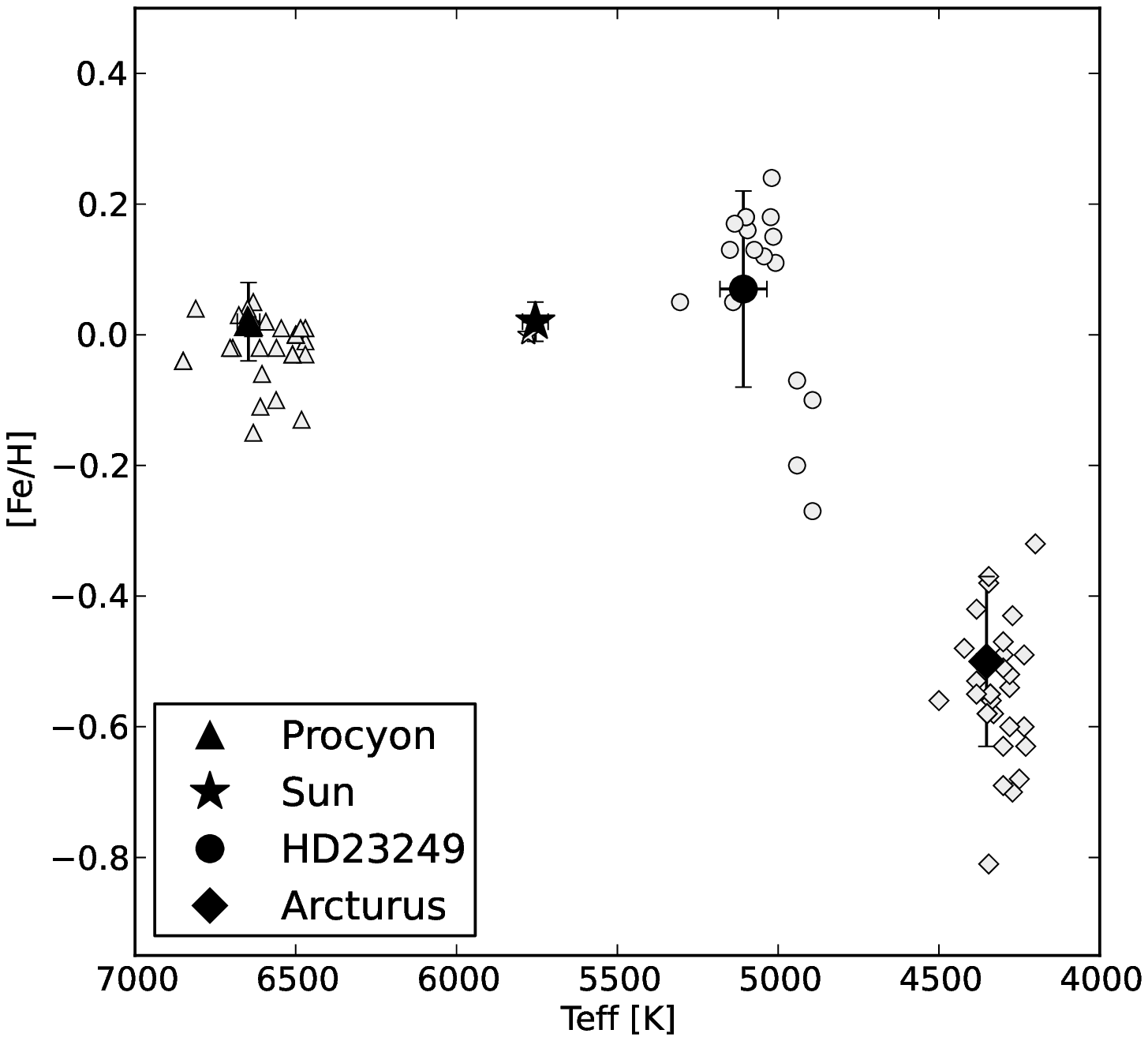}} \caption{\label{3bmTeffFeH}As in previous figure, but for T$_{\text{eff}}$ and [Fe/H].} \end{center}
\end{figure}

The effective temperatures, surface gravities and metallicities we obtain for these four stars are in good agreement with the values available in literature, as shown in Fig.~\ref{3bmTefflogg} and \ref{3bmTeffFeH}, and summarised in Table \ref{benchmarks_summary}.

\begin{table}
	\caption{ \label{benchmarks_summary} Atmospheric parameters of the four benchmark stars.}
	\small\addtolength{\tabcolsep}{-1pt}
	\begin{tabular}{l l l l l l l}
	  \hline
	  \hline
	  star & $\text{T}_{\text{eff}}$ & $\Delta\text{T}_{\text{eff}}$ & $\text{log}\,g$ & $\Delta\text{log}\,g$ & [Fe/H] & $\Delta$[Fe/H] \\
	       & [K] & [K] & & & & \\
	  \hline
          \multicolumn{7}{c}{Literature} \\
	  \hline
	  Arcturus & 4302 & 120 & 1.68 & 0.31 & -0.53 & 0.12  \\  
	  HD~23249 & 5025  & 255 & 3.84 & 0.17 & 0.07  & 0.15  \\  
	  Procyon & 6583 & 162 & 4.06 & 0.15 & -0.01 & 0.16  \\
	  Sun  & 5777 & ... & 4.44 & ... &  0 & ... \\
	  \hline
          \multicolumn{7}{c}{This study} \\
	  \hline
	  Arcturus & 4352 & 31  &   1.78 & 0.12  & -0.45 & 0.13  \\ 
	  HD~23249 & 5108 & 73  &   3.82 & 0.15  &  0.07 & 0.15    \\
	  Procyon & 6647 & 35  &   3.83 & 0.08  &  0.02 & 0.06     \\
	  Sun & 5755 & 40  &   4.30 & 0.20  &  0.02 & 0.09  \\
	  \hline
	\end{tabular}
\tablefoot{For the literature values the numbers are the average values found in PASTEL, and their associated errors are their standard deviations. For this study, the parameters and their errors are given by FAMA.}
\end{table}

\section[]{Conclusion}\label{sec_conclusion}
The current and future large-scale spectroscopic surveys require automatic procedures for batch-processing large numbers of stellar spectra. Based on DAOSPEC, DOOp provides a robust and convenient way of measuring EWs and produces results of the same quality as DAOSPEC used manually, while requiring less user intervention, thus making the results more reproducible and the process faster. 
DOOp is able to optimise the key parameters of DAOSPEC, but not the order of the polynomial used for the continuum fitting, which still has to be chosen by the user. We show that different choices of continuum order can lead to small differences in the EWs, up to $\pm$ 2\,m{\AA}, way within the errors in the measuring procedures.

\section*{Acknowledgements}
This work was partially supported by the Gaia Research for European Astronomy Training (GREAT-ITN) Marie Curie network, funded through the European Union Seventh Framework Programme (FP7/2007-2013) under grant agreement no. 264895.

\label{lastpage}


\begin{thebibliography}{99}
\bibitem[\protect\citeauthoryear{Allende Prieto et al.}{2008}]{apogee} Allende Prieto, C., Majewski, S. R., Schiavon, R., et al. 2008, Astronomische Nachrichten, 329, 1018
\bibitem[\protect\citeauthoryear{Bagnulo et al.}{2003}]{bagnulo} Bagnulo, S., Jehin E.~C., Cabanac, R., et al., ESO Paranal Science Operations Team, 2003, Msngr, 114, 10
\bibitem[\protect\citeauthoryear{Barden et al.}{2010}]{galah} Barden, S. C., Jones, D. J., Barnes, S. I., et al. 2010, Proc. SPIE, 7735
\bibitem[\protect\citeauthoryear{Bragaglia et al.}{2008}]{bragaglia08} Bragaglia, A., Sestito, P., Villanova, S., et al., 2008, A\&A, 480, 79
\bibitem[\protect\citeauthoryear{Cirasuolo et al.}{2011}]{moons} Cirasuolo, M., Afonso, J., Bender, R., et al. 2011, The Messenger, 145, 11
\bibitem[\protect\citeauthoryear{Dawson \& Ianna}{1998}]{Dawson98} Dawson, D.~W., Ianna, P.~A. 1998, AJ, 115, 1076
\bibitem[\protect\citeauthoryear{de Jong}{2011}]{4most} de Jong, R. 2011, The Messenger, 145, 14
\bibitem[\protect\citeauthoryear{Fitzpatrick \& Sneden}{1987}]{fitzpatrick87} Fitzpatrick, M.~J., Sneden, C., 1987, BAAS, 19, 1129
\bibitem[\protect\citeauthoryear{Friel et al.}{2003}]{friel03} Friel, E.~D., Jacobson, H.~R., Barrett, E., et al., 2003, AJ, 126, 2372
\bibitem[\protect\citeauthoryear{Gilmore et al.}{2012}]{ges12} Gilmore, G., Randich, S., Asplund, et al., 2012, Msngr, 147, 25
\bibitem[\protect\citeauthoryear{Gustafsson et al.}{2008}]{gust08} Gustafsson, B., Edvardsson, B., Eriksson, K., et al., 2008, A\&A, 486, 951 
\bibitem[\protect\citeauthoryear{Heiter et al.}{in prep.}]{Heiter13} Heiter, U., and the GES Line List group, in preparation.
\bibitem[\protect\citeauthoryear{Hinkle et al.}{2000}]{hinkle00} Hinkle, K., Wallace L., Valenti, J., et al., 2000, Visible and Near Infrared Atlas of the Arcturus Spectrum 3727-9300 A
\bibitem[\protect\citeauthoryear{Kurucz}{2005}]{kurucz05} Kurucz R.~L., 2005, MSAIS, 8, 14 
\bibitem[\protect\citeauthoryear{Magrini et al.}{2013}]{fama} Magrini, L., Randich, S., Friel, E.~D. et al., 2013, A\&A in press, astro-ph:1307.2367
\bibitem[\protect\citeauthoryear{Mucciarelli et al.}{2013}]{gala} Mucciarelli, A., Pancino, E., Lovisi, L., et al., 2013, ApJ, 766, 78
\bibitem[\protect\citeauthoryear{Pancino et al.}{2010}]{pancino10} Pancino, E., Carrera, R., Rossetti,  et al., 2010, A\&A, 511, 56
\bibitem[\protect\citeauthoryear{Ram{\'{\i}}rez et al.}{2001}]{ewdet} Ram{\'{\i}}rez, S. V., Cohen, J. G., Buss, J., Briley, M. M., 2001, AJ, 122, 1429
\bibitem[\protect\citeauthoryear{Recio-Blanco et al.}{2006}]{matisse} Recio-Blanco, A., Bijaoui, A., de Laverny, P., 2006, MNRAS, 370, 141
\bibitem[\protect\citeauthoryear{Sestito et al.}{2008}]{sest08} Sestito, P., Bragaglia, A., Randich, S., et al., 2008, A\&A, 488, 943 
\bibitem[\protect\citeauthoryear{Sneden et al.}{2012}]{sneden12} Sneden, C., Bean, J., Ivans, I., et al., 2012, Astrophysics Source Code Library, record ascl:1202.009 
\bibitem[\protect\citeauthoryear{Sousa et al.}{2007}]{ares} Sousa, S. G., Santos, N. C., Israelian, G., Mayor, M., Monteiro, M. J. P. F. G. 2007, A\&A, 469, 783
\bibitem[\protect\citeauthoryear{Stetson \& Pancino}{2008}]{dao08} Stetson, P.~B., Pancino, E., 2008, PASP, 120, 1332
\bibitem[\protect\citeauthoryear{Valentin \& Piskunov}{1996}]{sme} Valenti, J. A., Piskunov, N., 1996, A\&AS, 118, 595
\bibitem[\protect\citeauthoryear{Zwitter}{2008}]{rave} Zwitter, T., Siebert, A., Munari, U., et al. 2008, AJ, 136, 421
\end{thebibliography}
\end{document}